\documentclass[%
reprint,
amsmath,amssymb,
aps,
pra,
longbibliography,
]{revtex4-1}
\usepackage{graphicx}
\usepackage{dcolumn}
\usepackage{bm}

\usepackage{amssymb,amsmath}
\usepackage{lineno,hyperref}
\usepackage{placeins}

\begin{document}
	\title{Machine learning quantum mechanics: solving quantum mechanics problems using radial basis function network}
	\author{Peiyuan Teng}
	\email{teng.73@osu.edu}
	\affiliation{
		Department of Physics\\
		The Ohio State University\\
		Columbus, Ohio, 43210, USA
	}

\begin{abstract}
In this article, machine learning methods are used to solve quantum mechanics problems.  The radial basis function(RBF) network in a discrete basis is used as the variational wavefunction for the ground state of a quantum system. Variational Monte Carlo(VMC) calculations are carried out for some simple Hamiltonians. The results are in good agreements with theoretical values. The smallest eigenvalue of a Hermitian matrix can also be acquired using VMC calculations. New results are provided to demonstrate that machine learning techniques are capable of solving quantum mechanical problems.
\end{abstract}

\maketitle

\section{Introduction}

Machine learning theory has been developing rapidly in recent years. Machine learning techniques have been successfully applied to solve a variety of problems, such as email filtering, optical character recognition(OCR), and natural language processing, and have become a part of everyday life. In the physical sciences, researchers are also applying machine learning methods to explore new possibilities. For example, machine learning methods are used in molecular dynamics\cite{C5SC04786B}\cite{C7SC02267K}, as a way to bypass the Kohn-Sham equation in density functional theory\cite{Brockherde2017}, to assist in materials discovery\cite{Raccuglia2016}, or to identify phase transitions\cite{Carrasquilla2017}. Considering the power of machine learning, it is interesting to consider solving quantum mechanics problems using machine learning methods.

Artificial neural networks (ANNs) \cite{McCulloch1943}, which are inspired by biological neural networks, are one of the most important methods in machine learning theory. An ANN consists of a network of artificial neurons, and examples of ANNs include feedforward neural networks\cite{bishop2006pattern}, radial basis function (RBF) networks\cite{SCHWENKER2001439}, and restricted Boltzmann machines\cite{Hinton504}.
As a universal approximator \cite{SCARSELLI199815}\cite{doi:10.1162/neco.1991.3.2.246}, an ANN can be used to represent functions, and it is possible to use an ANN as a representation of the wavefunction in a quantum system. 

Researchers have been trying to combine neural network theory and quantum mechanics, for example, using a neural network in the real space to solve differential equations, especially the schr\"{o}dinger equation with some specific potential\cite{LAGARIS19971}. Another example is the quantum neural network\cite{DASILVA201655}, where information in an ANN is processed quantum mechanically. One of the most promising works was the recent research by Carleo and Troyer in Ref.\cite{Carleo602}, where the restricted Boltzmann machine was used as the variational Monte Carlo (VMC) ground state wavefunction. In their work, the ground state of a many-body system could be efficiently represented by a neural network. Following their work, other possibilities were also explored. Most recently, in Ref.\cite{doi:10.7566/JPSJ.86.093001}, a three-layer feedforward neural network was used to calculate the ground state energy of the Bose-Hubbard model. Machine learning methods were shown to be able to distinguish between different phases, even for systems with the sign problem\cite{sign}. VMC methods do not suffer from the fermion sign problem; therefore, using a neural network as a VMC ansatz is very promising and has the potential to tackle the calculations that are almost impossible in other Monte Carlo methods.

In this article, the possibility of using an RBF network to represent the wavefunction of a quantum-mechanical system is discussed. Our work is new in two major aspects. First, the representation power of the RBF network is illustrated, which has not been discussed in the physics literature. Second, instead of a lattice system, where the dimension of the Hilbert space of each site is finite, a general quantum-mechanical system with infinite or continuous degrees of freedom is discussed. A binary restricted Boltzmann machine is not sufficient for the simulations of such a system; therefore, it is interesting to search for new ansatz. An RBF network is one of the candidates.

In our work, a VMC procedure is formulated, where an RBF network is used as the variational wavefunction. A harmonic oscillator in a linear potential and a particle in a box with a linear potential are then used as benchmarks.  Furthermore, we discuss the possibility of using the VMC method to solve for the lowest eigenvalue of a matrix.

This article is organized as follows. In section \ref{sec2}, artificial neural network theory and variational Monte Carlo theory are reviewed. Section \ref{sec3} contains major results, that is, quantum mechanical problems are solved using the radial basis neural network. In section \ref{sec4}, we discuss some related questions.

\section{Artificial neural network theory and the variational Monte Carlo method} \label{sec2}
In this section, two cornerstones of this work will be introduced, which are the artificial neural network theory and the variational Monte Carlo method. 

\subsection{Artificial neural network theory}
\begin{figure}[!htb]
	\includegraphics[width=65mm]{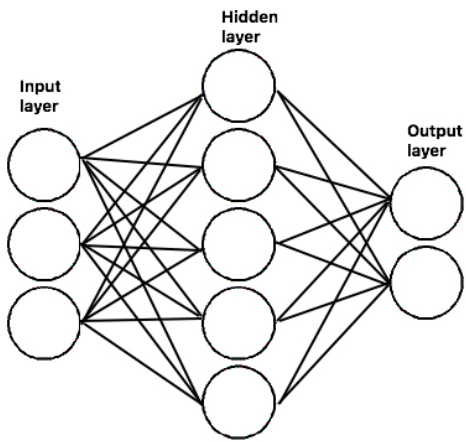}
	\caption{An illustration of the artificial neural network. A typical neural network consists of three layers of neurons: the input layer, the hidden layer, and the output layer. Each neuron is represented by a circle.  The lines between layers are associated with the parameters of the neural network.}
\label{fig:fig1}
\end{figure}

Inspired by the biological neural network model, ANN theory was proposed by McCulloch and Pitts in 1943\cite{McCulloch1943}, in an attempt to propose a mathematical description of the biological nervous system.  Figure. \ref{fig:fig1}  illustrates a simple example of a neural network which consists of three layers of artificial neurons. 

Neural networks are widely-used tools in machine learning theory, for example, as a function approximation tool in supervised learning. The goal is to find the optimal parameters by minimizing the cost function. This can be a highly non-trivial problem when there are a large number of parameters. For such algorithms such as the back-propagation,  please see Ref.\cite{bishop2006pattern}.

In a typical machine learning problem using neural network methods, the input neuron can be a binary number. For example, in a handwritten digit recognition problem, each input neuron corresponds to a pixel in a figure and takes a value of 0 or 1. The input values are processed through the neural network using, for example, the rules mentioned above. The output values of the neural network are compared with the objective values, and the error  is minimized by finding the optimal parameters.

In this article, the radial basis function (RBF) network, is used as a variational wavefunction ansatz.   For example, for a three-layer RBF network with one single output neuron, the output function $z(\boldsymbol{x})$ of the neural network can be written as

 \begin{equation}
z(\boldsymbol{x})=\sum_{i=1}^M a_i \rho_i(||\boldsymbol{x}-\boldsymbol{c}_i||).
\end{equation}

In this output function, $a_i$ and $\boldsymbol{c}_i$ are parameters of the neural network. $\boldsymbol{x}$ is the input vector which has the same dimension as $\boldsymbol{c}_i$. $M$ is the number of neurons in the hidden layer. $\rho(||\bullet||)$ is the radial basis function which can be a Gaussian function with a Euclidean norm.
 \begin{equation}  \label{Gaussian}
\rho_i(||\boldsymbol{x}-\boldsymbol{c}_i||)=e^{-|b_i||\boldsymbol{x}-\boldsymbol{c}_i|^2},
\end{equation}
 or an exponential absolute value function
 
  \begin{equation} \label{abs}
\rho_i(||\boldsymbol{x}-\boldsymbol{c}_i||)=e^{-|b_i||\boldsymbol{x}-\boldsymbol{c}_i|}.
\end{equation}

Other activation functions, such as multiquadratics 

  \begin{equation} 
\rho_i(||\boldsymbol{x}-\boldsymbol{c}_i||)=\sqrt{|\boldsymbol{x}-\boldsymbol{c}_i|^2-|b_i|^2},
\end{equation}

Or inverse multiquadratics

  \begin{equation} 
\rho_i(||\boldsymbol{x}-\boldsymbol{c}_i||)=({|\boldsymbol{x}-\boldsymbol{c}_i|^2-|b_i|^2)^{-\frac{1}{2}}},
\end{equation}
 are also commonly used in the machine learning community. These activation functions can also be understood as kernel functions. In the activation functions, $|b_i|$ are parameters that control the spread of the activation function. Other activation function are also possible, discussions about the activation function can be found in Ref. \cite{58326}
 
In addition to the RBF network, many different types of neural networks can be constructed, such as the restricted Boltzmann machines or the autoencoders, which are widely used in deep learning technology. The universal approximation theorem establishes the mathematical foundation of neural network theory, which states that neural network functions are dense in the space of continuous functions defined on a compact subset of $R^n$, under some assumptions about the activation function and given enough hidden neurons\cite{SCARSELLI199815}\cite{doi:10.1162/neco.1991.3.2.246}.

In this paper, the RBF network is used as a variational wave function represented in a discrete eigenbasis. Note that we use $|b_i|$ as a variational parameter in our calculations instead of a constant number as in a regular RBF network. The absolute value of $|b_i|$ is for the stability of the optimization.

When neural network methods are applied to quantum physics, the inputs of the neural network can take discrete quantum numbers.  After being processed through the neural network, the outputs of the neural network represent the amplitudes of the wavefunction on the basis labeled by the input quantum numbers. The neural network is then trained by minimizing the energy expectation value. For example, for a three-dimensional  quantum harmonic oscillator in an orthogonal coordinate system, we can use a neural network with three input neurons, where each input can take integer values for 0 to $\infty$. The trained neural network should represent the ground state of this system, in which, after proper normalization, the output should be $1$ given $000$ as the input, and $0$ for other inputs.

\subsection{Variational Monte Carlo method(VMC)}

The VMC method, first proposed by McMillan in 1965\cite{PhysRev.138.A442}, combines the variational method and the Monte Carlo method in order to evaluate the ground state of a quantum system.

Start from a Hamiltonian $\hat{H}$ and a variational wave function $|\psi(\lambda)\rangle$, where $\lambda$ is a set of variational parameters, the energy expectation value can be written as

 \begin{equation}
E(\lambda)=\dfrac{\langle\psi(\lambda)|\hat{H}|\psi(\lambda)\rangle}{\langle\psi(\lambda)|\psi(\lambda)\rangle}.
\end{equation}

This energy expectation value can be computed using the widely known Metropolis algorithm\cite{doi:10.1063/1.1699114}, which is one of the most efficient algorithms in computational science. As a Markov chain Monte Carlo method, it may currently be the only efficient algorithm for evaluating a multidimensional integral.

The next step of the VMC method is to minimize the energy in the parameter space. This can be a difficult problem when there are many variational parameters. Two examples of such algorithms are the linear method\cite{PhysRevLett.98.110201} and the stochastic reconfiguration method\cite{PhysRevB.71.241103}. The minimization algorithm gives the minimum of the energy in the parameter space, and it is reasonable to use this value as our approximation for the ground state energy. For a detailed review of the VMC method, please refer to Ref. \cite{Rubenstein2017}. 

Currently, physicists believe that the accuracy of the VMC method depends, to a great extent, on a proper choice of the variational wavefunction; therefore, it is important to choose a wavefunction based on physical intuition or a physical understanding of the system. This belief may not be true in the age of machine learning. Neural network functions are capable of approximating unknown functions by maximizing or minimizing the objective function. It would be interesting to further explore the possibility of using a neural network function as the variational wavefunction of a quantum system.

\section{Solving quantum mechanics problems using artificial neural network} \label{sec3}

In the pioneering work of Carleo and Troyer\cite{Carleo602}, restricted Boltzmann machine(RBM) was used as a variational wave-function for many-body systems. The transverse-field Ising model and anti-ferromagnetic Heisenberg model were benchmarked using the RBM wavefunction. Variational Monte Carlo calculations were carried out. Their results demonstrate that a neural network wavefunction is capable of capturing the quantum entanglement of the ground states and giving an accurate estimation of the ground state energy.

In this article, we continue developing this idea using artificial neural network functions as the ground state variational wavefunction. In Ref.\cite{Carleo602}, the restricted Boltzmann machine is only binary-valued, we will demonstrate the representation power of a neural network wavefunction without this constraint. In addition, we discuss the possibility of using a neural network wavefunction to solve a generic quantum mechanics problem. This VMC method behaves at least as accurate as the perturbation theory.

\subsection{Theoretical outline} \label{outline}

Consider a quantum system which has countable number of basis, an arbitrary state $|\psi\rangle$ in the Hilbert space can be represented by 
 
 \begin{equation}
|\psi\rangle=\sum_{n_1,n_2,...,n_p}\psi(n_1,n_2,...,n_p)|n_1,n_2,...,n_p\rangle,
\end{equation}

where $|n_1,n_2,...,n_p\rangle$ is a set of basis labeled by quantum number $n_i$, $i=1,2...p.$, and $p$ is the number of sites in the system. For example, for the Heisenberg model, $p$ represents the number of spins; for a three dimensional harmonic oscillator in a Cartesian coordinate, we could use $n_1$,$n_2$,$n_3$ to label three quantum numbers. $\psi(n_1,n_2,...,n_p)$ is the amplitude of $|\psi\rangle$ on basis $|n_1,n_2,...,n_p\rangle$. We can interpret this amplitude as a function of $n_1,n_2,...,n_p$. A similar ansatz is also used in Ref. \cite{doi:10.7566/JPSJ.86.093001}.

This function can be represented by a neural network with \textit{one} output neuron. Using an RBF network, the amplitude function can be written as,

\begin{equation} 
\psi(n_1,n_2,...,n_p;\boldsymbol{a},\boldsymbol{c})=\sum_i^M a_i \rho_i(||\boldsymbol{n}-\boldsymbol{c}_i||),
\end{equation}

with $\boldsymbol{n}$ represents an array of quantum numbers and

 \begin{equation} 
\rho_i(||\boldsymbol{x}-\boldsymbol{c}_i||)=e^{-|b_i| |\boldsymbol{x}-\boldsymbol{c}_i|^2}.
\end{equation}

One reason to choose this neural network is that the Gaussian activation function guarantees that the amplitude does not diverge when $n\rightarrow \infty$.

Practically, it is useful to truncate the quantum number $n_i$ if its range is countably infinite. This is not necessary for a spin half lattice system since $n_i$ can only take two values. For a harmonic oscillator, however, we may truncate the quantum number at some finite value. The universal approximation theorem is only valid for a closed space. This truncation will also facilitate numerical simulations. 

Using this variational wave function, the energy expectation value is

 \begin{equation}
E(\boldsymbol\lambda)=\dfrac{\langle\psi(\boldsymbol\lambda)|H|\psi(\boldsymbol\lambda)\rangle}{\langle\psi(\boldsymbol\lambda)|\psi(\boldsymbol\lambda)\rangle}=\dfrac{\int |\psi(\boldsymbol{n};\boldsymbol\lambda)|^2 E_{local}(\boldsymbol{n};\boldsymbol\lambda))d\boldsymbol{n}}{\int |\psi(\boldsymbol{n};\boldsymbol\lambda)|^2d\boldsymbol{n}},
\end{equation}

 with
 
 \begin{equation}
E_{local}(\boldsymbol{n};\boldsymbol\lambda)= \dfrac{\langle\boldsymbol{n}|H|\psi(\boldsymbol\lambda)\rangle}{\langle\boldsymbol{n}|\psi(\boldsymbol\lambda)\rangle}= \dfrac{\sum_{n'}\langle\boldsymbol{n}|H|\boldsymbol{n'}\rangle\langle\boldsymbol{n'}|\psi(\boldsymbol\lambda)\rangle}{\langle\boldsymbol{n}|\psi(\boldsymbol\lambda)\rangle},
\end{equation}

Here, $\boldsymbol\lambda$ represents all the variational parameters, for example, $a_i$, $b_i$ and $\boldsymbol{c}_i$.

The energy expectation can be evaluated using the Metropolis algorithm. After initialization and thermalization, repeat these two step until equilibrium: (1) generate a move from configuration $\boldsymbol{n}$ to $\boldsymbol{n''}$. (2) Using proper transition probability, accept or reject the move with probability $min(1,\vert\frac{\langle\boldsymbol{n''}|\psi(\boldsymbol\lambda)\rangle}{\langle\boldsymbol{n}|\psi(\boldsymbol\lambda)\rangle}\vert^2)$. Expectation value of other operators can be evaluated similarly.

Compared with exact diagonalization, one advantage of this formalism is that the matrix element $\langle\boldsymbol{n}|H|\boldsymbol{n'}\rangle$ is never stored explicitly. Only the non-zero matrix elements are needed to be valued and summed during the sampling process.

The energy as a function of parameters $\boldsymbol\lambda$ can be, for example, minimized using the stochastic reconfiguration method\cite{PhysRevB.71.241103}.  In the stochastic reconfiguration method, an operator

 \begin{equation}
O_{i}(\boldsymbol{n})=\dfrac{\partial_{\lambda_i}\psi_{\boldsymbol\lambda}(\boldsymbol{n})}{\psi_{\boldsymbol\lambda}(\boldsymbol{n})},
\end{equation}

can be defined for each parameter in the variational wavefunction.

For a radial basis neural network with the Gaussian basis function

 \begin{equation}
O_{a_{i}}(\boldsymbol{n})=\dfrac{\rho_i}{\psi},
\end{equation}

 \begin{equation}
O_{b_{i}}(\boldsymbol{n})=-\dfrac{a_i b_i|\boldsymbol{n}-\boldsymbol{c}_i|^2\rho_i}{|b_i|\psi },
\end{equation}

 \begin{equation}
O_{c_{ij}}(\boldsymbol{n})=\dfrac{2 a_i |b_i|(n_j-c_{ij})\rho_i}{\psi },
\end{equation}

where $c_{ij}$ is the j-th component of of $\boldsymbol{c}_i$.
The covariance matrix and forces are defined as

 \begin{equation}
S_{ij}=\langle O^{*}_{i}O_{j} \rangle-\langle O^{*}_{i}\rangle\langle O_{j} \rangle,
\end{equation}

 \begin{equation}
F_{i}=\langle E_{local}O^{*}_{i} \rangle-\langle E_{local} \rangle\langle O^{*}_{i}\rangle.
\end{equation}

The parameters can be updated by

 \begin{equation}
 \lambda'_j= \lambda_j+\alpha S^{-1}_{ij}F_{i}.
\end{equation}

Here, $\langle\bullet \rangle$ is the expectation value of an operator. $\alpha$ can be understood as the learning rate of the optimization algorithm. A regularization, $S'_{ii}=S_{ii}+r(k)S_{ii}$, is applied to the diagonal elements of matrix $S$ in all our calculation, where $r(k)=max(100\times0.9^k,10^{-4})$\cite{Carleo602}. This process iterates until the optimization converges, and we treat the converged energy as our best approximation of the ground state energy.

In this article, the method mentioned above is used for the optimization. We notice that the recent work of Saito\cite{doi:10.7566/JPSJ.86.093001}, in which feedforward neural network was successfully used to represent the ground state of the Bose-Hubbard model. In their work, an exponential function was written based on the output of the feedforward neural network. It is an interesting question whether an exponential of feedforward neural network output function can be used to represent a quantum mechanical wavefunction.

\subsection{One dimensional quantum harmonic oscillator in electric field}

To start with, we'd like to benchmark the quantum harmonic oscillator.€ Since we use a set of discrete quantum numbers to describe the variational wavefunction, it is natural to use the energy eigenbasis of an unperturbed harmonic oscillator to calculate the matrix element.

Consider the one dimensional Hamiltonian
 \begin{equation}
H=\dfrac{\hat{p}^2}{2}+\dfrac{\hat{x}^2}{2}+E\hat{x}=H_0+E\hat{x},
\end{equation}

where $E$ is a parameter that can be understood as the electric field.

Using natural units, it is easy to see that the ground state energy of $H_0$ is $0.5$.

Assuming the eigenstates of $H_0$ are labeled by $|n\rangle$, the variational ansatz for the ground state of $H$ can be approximated by

 \begin{equation}
|\psi\rangle=\sum_{n=0}^{n_{max}-1}\psi(n)|n\rangle,
\end{equation}

with $\psi(n)$ represented by an RBF network with one input neuron, and we truncate the quantum number to $n_{max}-1$. In this notation, the RBF network represents the function $\psi(n)$. The variable $n$ can take different values, for example, if $n=1$, the output of the neural network is the coefficient on the basis $|1\rangle$, which is $\psi(1)$. The neural network represents the function $\psi$, and the coefficient on basis $|n \rangle$ are represented by $\psi(n)$.

We use the VMC procedure described in Section \ref{outline} to conduct the calculation. The parameters are initialized randomly.  Our codes are written in C++, where the matrix solving library Eigen\cite{eigenweb} is used for the Stochastic Reconfiguration. Sample codes will be available at https://github.com/peiyuanteng.

A neural network with random parameters is first created. And then the ground state energy under one set of parameters are calculated using the Monte Carlo method. The state space of the Monte Carlo sampling is a truncated discrete space denoted by $\boldsymbol{n}$. Specifically, our quantum number is the quantum number of the unperturbed Hamiltonian $H_0$, and the basis are the eigenbasis of $H_0$. We are trying to solve for the ground state of the perturbed one. A random plus or minus move is generated for each sample and accepted using the Metropolis Algorithm. In this work, when a random move yields a quantum number that is below zero or above $n_{max}-1$ at the boundary of state space, the quantum number is reflected back in order to satisfy the detailed balance condition.  For each specific $\boldsymbol{n}$, we can plug it into the neural network and get its amplitude. During the Monte Carlo process, 50000 samples are used. Being able to calculate the energy, we can then use the Stochastic Reconfiguration method to find the minimal energy, and we treat this energy as our best approximation of the ground state energy.

\begin{figure}[!htb]
	\includegraphics[width=90mm]{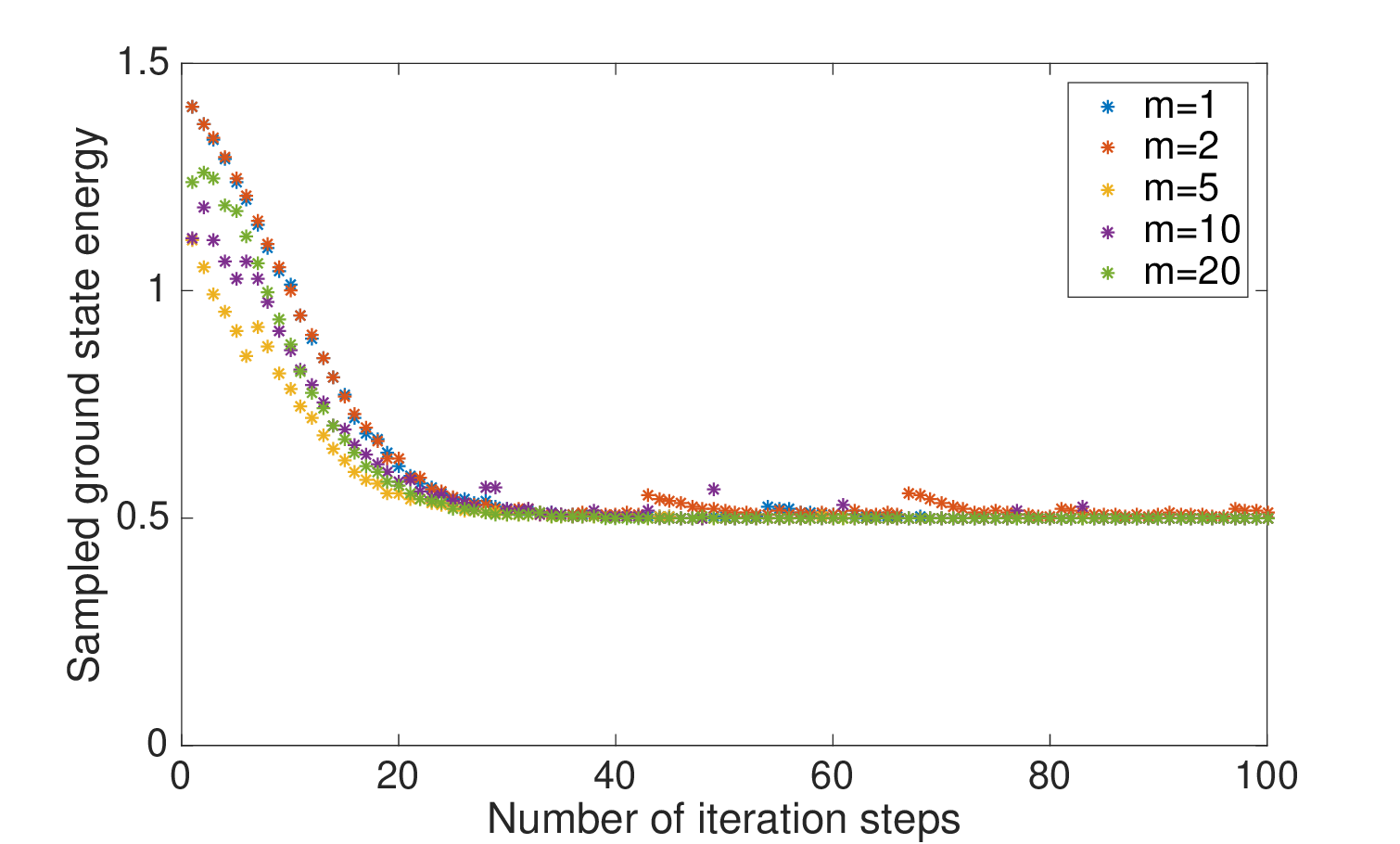}
	\caption{Minimization of the ground state energy of $H$ at $E=0$, using Gaussian radial basis network. m is the number of hidden layers in the neural network.}
\label{fig:fig2}
\end{figure}

In Figure. \ref{fig:fig2}, we illustrate the minimization of ground state energy during the iteration process using the Gaussian basis function. See Eq.\ref{Gaussian}. The learning rate is set at $0.1$. $m$ denotes the number of neurons in the hidden layer.

\begin{figure}[!htb]
	\includegraphics[width=90mm]{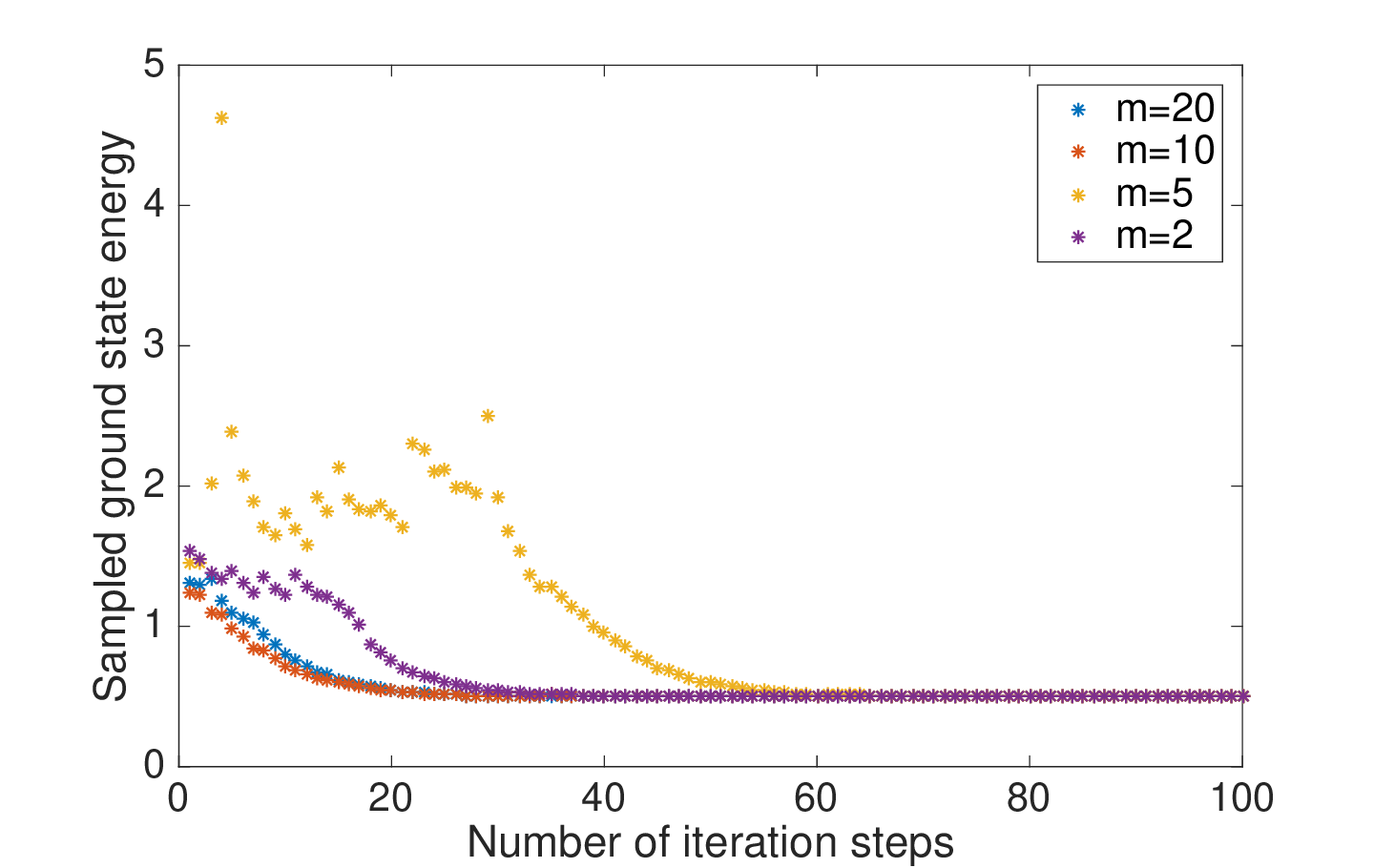}
	\caption{Minimization of the ground state energy of $H$ at $E=0$, using Eq. \ref{abs} as the radial basis function. m is the number of hidden layers in the neural network.}
\label{fig:fig3}
\end{figure}

Alternatively, we can use the exponential absolute value function as the RBF, see Eq. \ref{abs}.  Under the same learning rates, this RBF network also converges to the correct eigenvalue, see Figure.\ref{fig:fig3}. It is easy to see that the Gaussian RBF network behaves better than the other. Based on our experience, the Gaussian network also performs better in other cases, therefore we use the Gaussian network in later examples.

\textbf{Remarks:} We use $n$ as our variable for the variational wavefunction. The output of $\psi(n)$ is discrete. It should not be confused with the method that uses a Gaussian function in the coordinate representation as the variational wavefunction, which is trivial. One reason that we compare Eq. \ref{Gaussian} and Eq. \ref{abs} is to demonstrate that this method is capable of giving the correct coefficients regardless of the radial basis function.

\begin{figure}[!htb]
	\includegraphics[width=90mm]{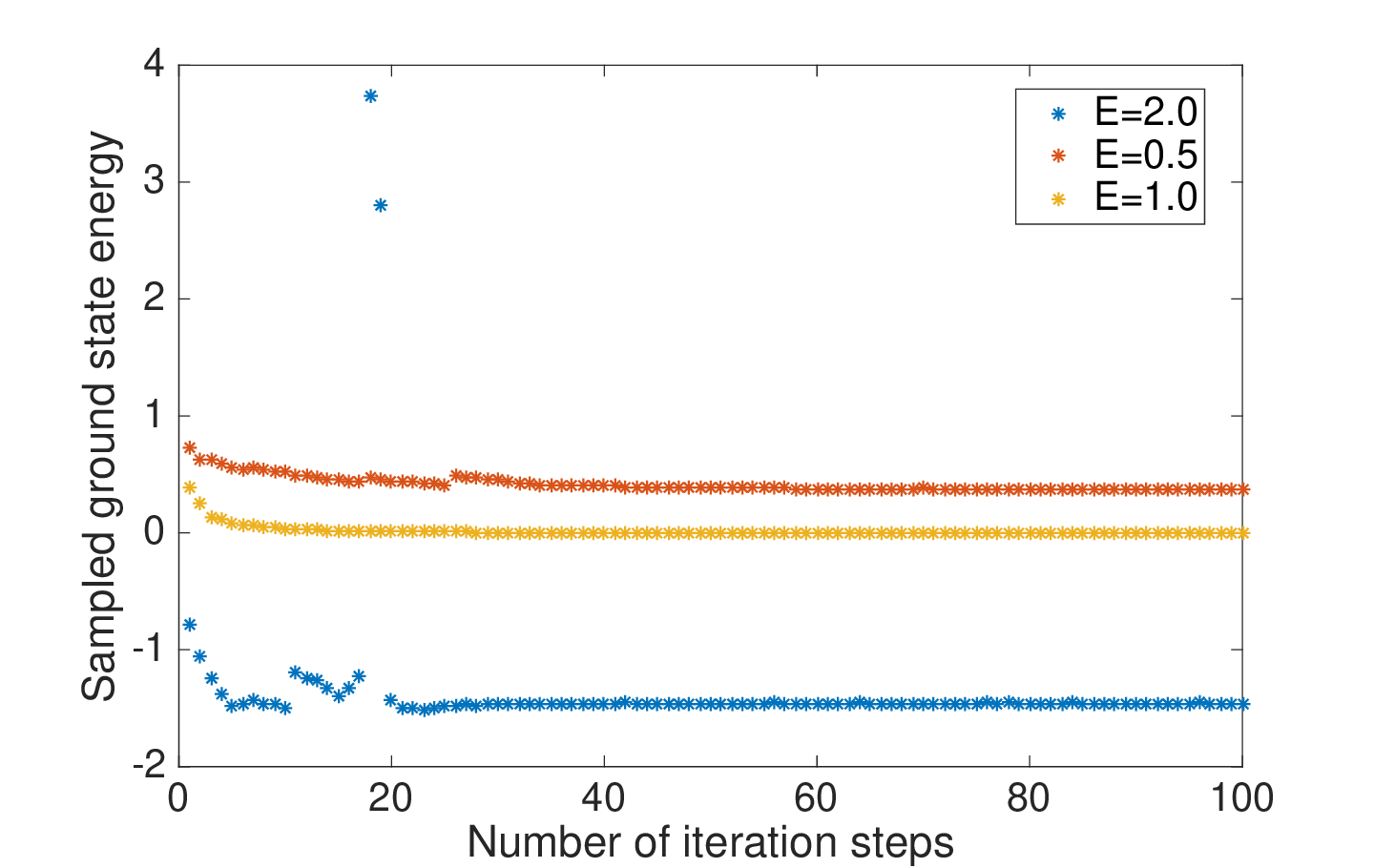}
	\caption{Minimization of the ground state energy of $H$ at $E=0.5,1.0,2.0$, using Gaussian radial basis function.}
\label{fig:fig4}
\end{figure}

Figure. \ref{fig:fig4}  illustrates the behavior of VMC under different electric field.  In our simulation, a separate neural network is trained for each $E$. The theoretical value of the ground state energy $e_g$ is $e_g=0.5(1-E^2)$. The VMC results converge at $0.375\pm0.000$, $0.000\pm0.000$,$-1.446\pm0.003$ when $E=0.5,1.0,2.0$, while the exact value is at $0.375$, $0.000$, $-1.5$ respectively. Notice that the error increase with $E$ under certain $nmax$. In this section $nmax=20$.  Expetation value and errors in this article are calculated when optimization is saturated.

Notice that during the optimization process, the sampled ground state energy may have some spikes. The author believes that this phenomenon is a result of the stochastic nature of the optimization algorithm. Random fluctuations of the expectation value of the operator and the complicated structure of the energy function may lead to drastic changes in the ground state energy during the optimization process.

\begin{figure}[!htb]
	\includegraphics[width=90mm]{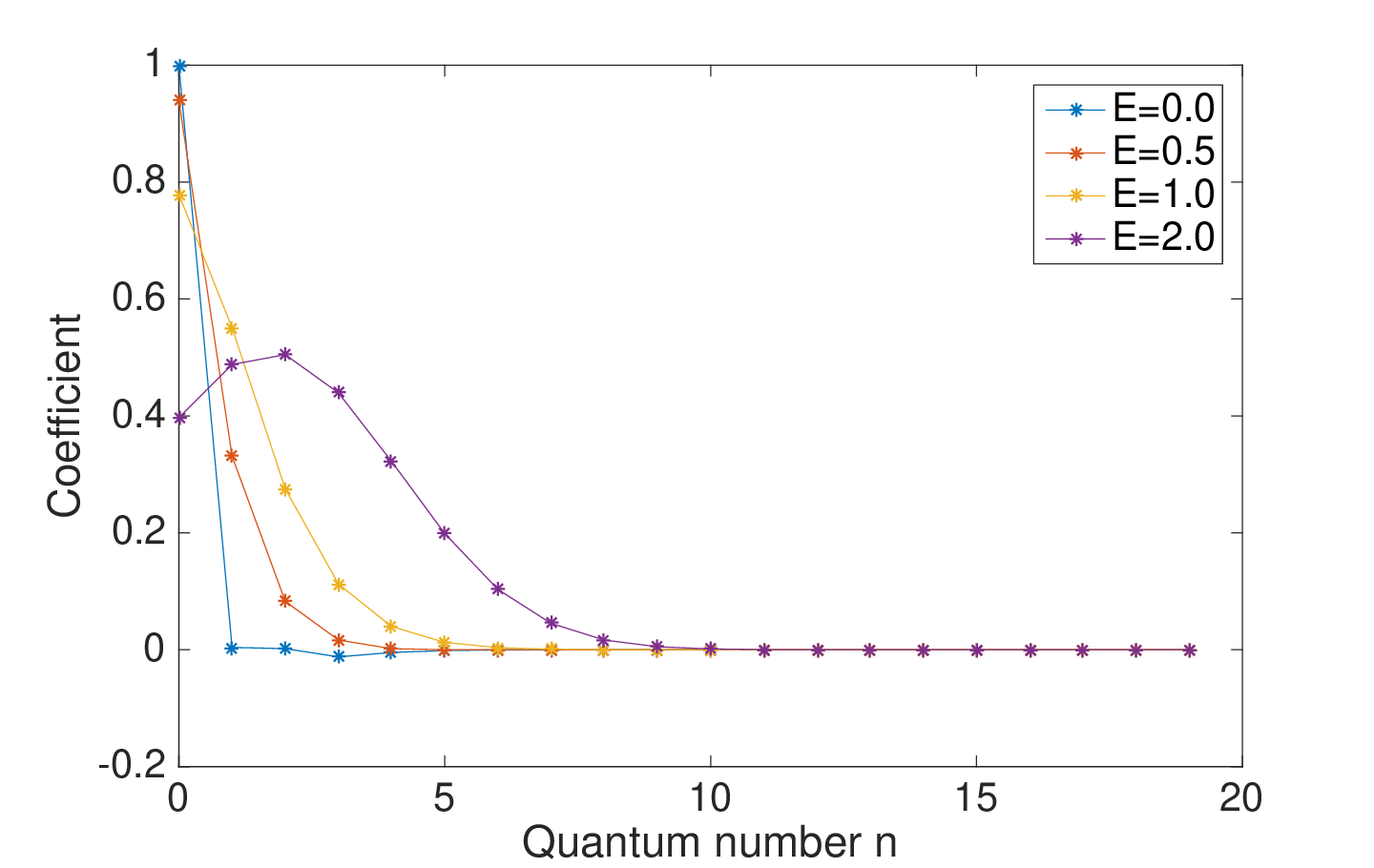}
	\caption{$\psi(n)$ as a function of $n$ at $E=0.0,0.5,1.0,2.0$, using Gaussian radial basis function. Circles represent theoretical values and asterisk represents the values with RBF network.}
\label{fig:fig5}
\end{figure}

Figure. \ref{fig:fig5} shows  $\psi(n)$ as a function of $n$ under different $E$. $\psi(n)$ is normalized and its value means the overlap between new ground state of $H$ and the energy eigenstate $|n\rangle$ of $H_o$.

Theoretically one can calculate that

 \begin{equation}
\psi(n)=\int_{-\infty}^{\infty}\dfrac{1}{\sqrt{2^n n!}} (\dfrac{1}{\pi})^{\frac{1}{2}}e^{-(x-E)^2/2}e^{-x^2/2}H_n(x)dx,
\end{equation}

where $H_n(x)$ are the Hermite polynomials. Simplify this expression, we will get

 \begin{equation}
\psi(n)=\dfrac{1}{\sqrt{2^n n!}}E^n e^{-E^2/4}.
\end{equation}

It can be seen that VMC values agree very well with the exact value when E is small. Errors begin to increase when E gets larger.

Based on these results, we claim that the radial basis neural network clearly captures the behavior of the 1D quantum harmonic oscillator.

\subsection{Two dimensional quantum harmonic oscillator in  electric field}

Similarly, we can consider a radial basis neural network with many input neurons. For example, with two input neurons, we can consider a two-dimensional quantum harmonic oscillator in an electric field.

Consider a Hamiltonian
 \begin{equation}
 H=\dfrac{\hat{p_x}^2}{2}+\dfrac{\hat{x}^2}{2}+\dfrac{\hat{p_y}^2}{2}+\dfrac{\hat{y}^2}{2}+E_x\hat{x}+E_y\hat{y}=H_0+E_x\hat{x}+E_y\hat{y}.
 \end{equation}

It is easy to see that the eigenvalue of $H_0$ is $1.0$. We will treat $E_x$ and $E_y$ as our parameters.

Our neural network wavefunction can be written as
 \begin{equation}
|\psi\rangle=\sum_{n_x,n_y=0}^{n_{max}-1}\psi(n_x, n_y)|n_x, n_y\rangle
\end{equation}

We can use the same VMC procedure as the previous part to perform the calculation. The learning rate, in this case, is set at $0.2$, our neural network has $10$ hidden neurons and $2$ input neurons. The algorithm used for this 2d example is similar to the 1d Harmonic Oscillator.

\begin{figure}[!htb]
	\includegraphics[width=80mm]{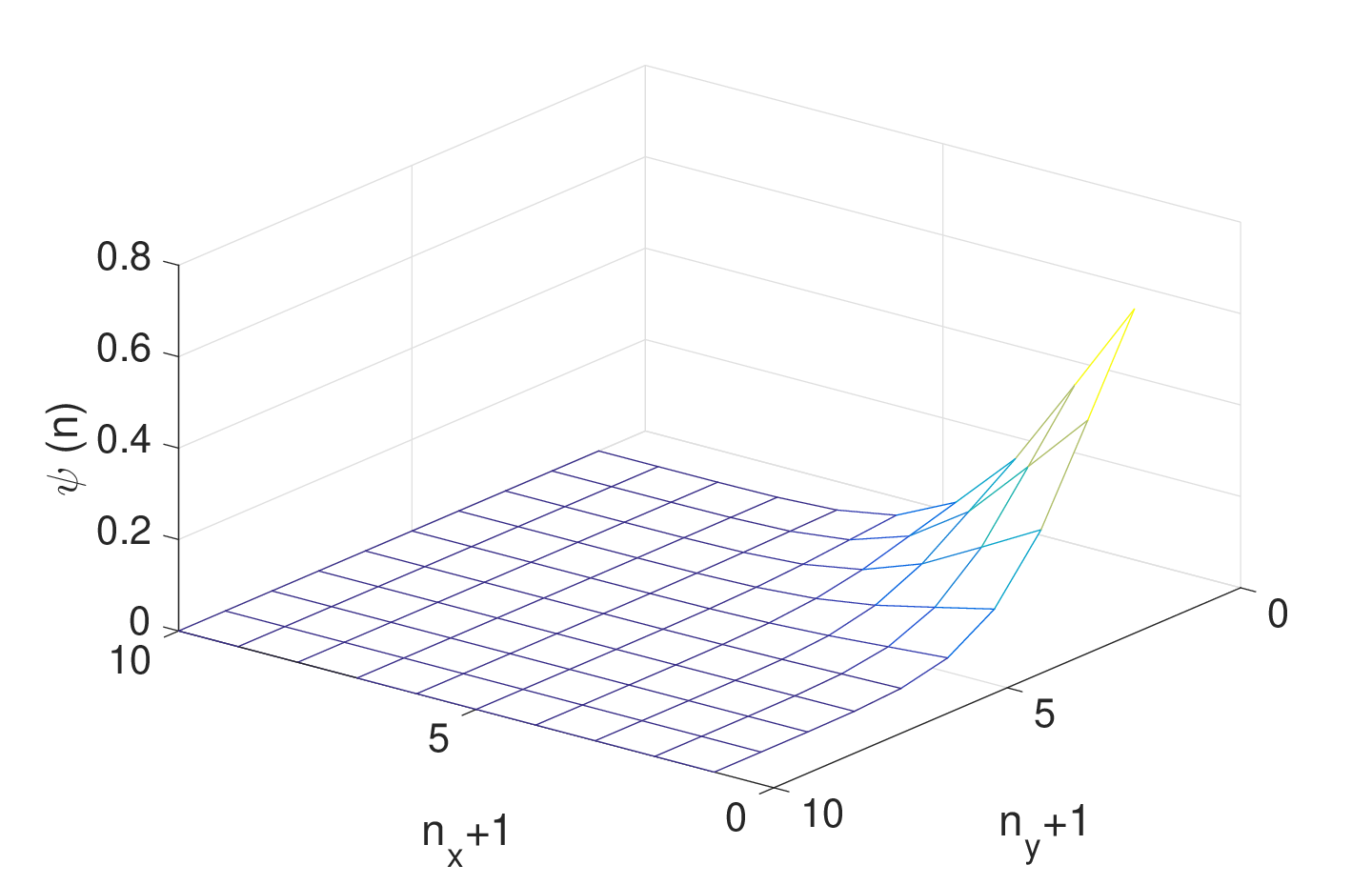}
	\caption{$\psi(n_x, n_y)$ as a function of $n_x+1, n_y+1$ at $E_x=1.0,E_y=1.0$, using Gaussian radial basis function. In this figure, $\psi(n_x, n_y)$  is not normalized.}
\label{fig:fig6}
\end{figure}

\begin{figure}[!htb]
	\includegraphics[width=80mm]{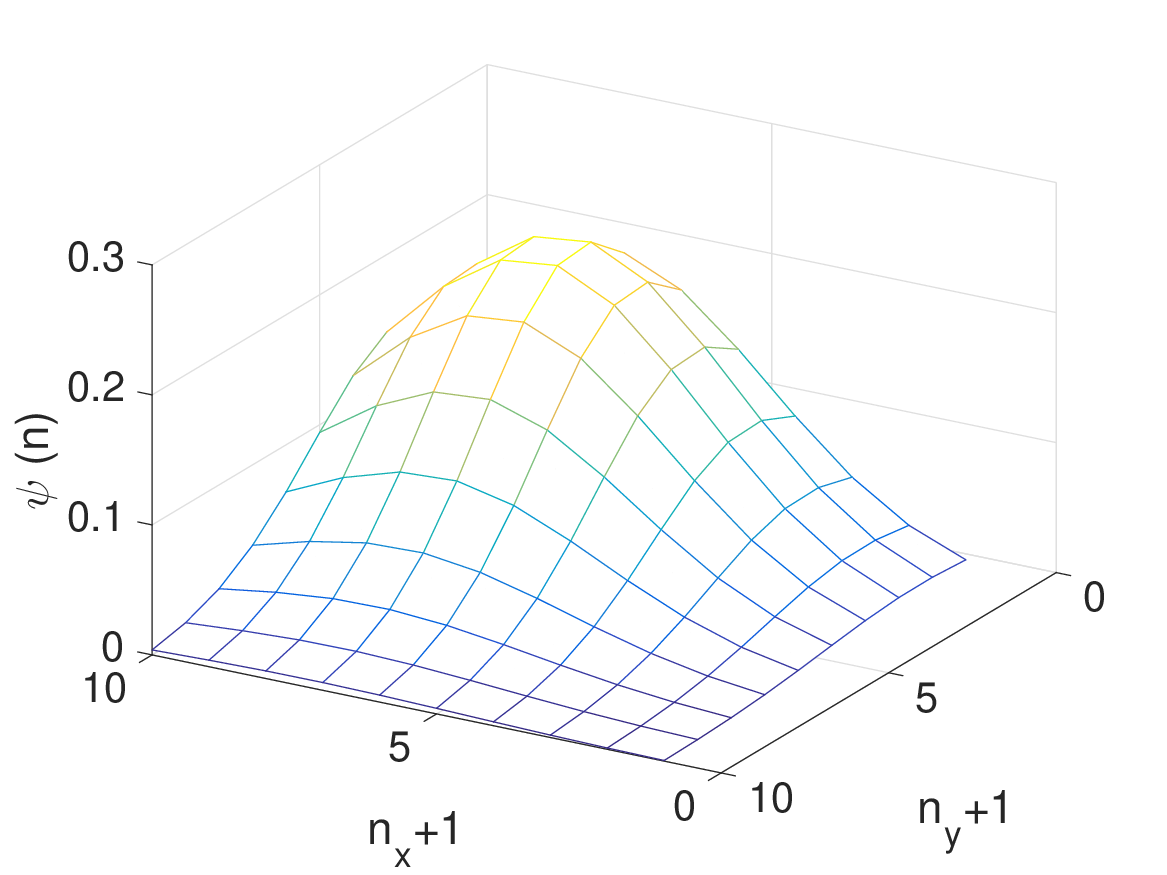}
	\caption{$\psi(n_x, n_y)$ as a function of $n_x+1, n_y+1$ at $E_x=4.0,E_y=2.0$, using Gaussian radial basis function.  In this figure, $\psi(n_x, n_y)$  is not normalized.}
\label{fig:fig7}
\end{figure}

Figure \ref{fig:fig6} and \ref{fig:fig7} illustrate the behavior of the trained neural network at different electric field. From the shape of the surface, we can see that a proper choice of $nmax$ is important to the accuracy of this method. The reason is that, in this example, when $E_x$, and $E_y$ gets larger, the bump in the function $\psi(n)$ will shift away from the origin. The states out of $nmax$ are not considered, therefore the accuracy will be affected if the overlaps out of $nmax$ are large. In these figures, we choose $nmax=10$ to illustrate the influence of $nmax$ on the accuracy.

The exact value of $\psi(n_x,n_y)$ can be solved as

 \begin{equation}
\psi(n_x,n_y)=\dfrac{1}{\sqrt{2^n_x n_x!}}\dfrac{1}{\sqrt{2^n_y n_y!}}E_x^{n_x} E_y^{n_y}e^{-E_x^2/4}e^{-E_y^2/4}.
\end{equation}

\begin{figure}[!htb]
	\includegraphics[width=80mm]{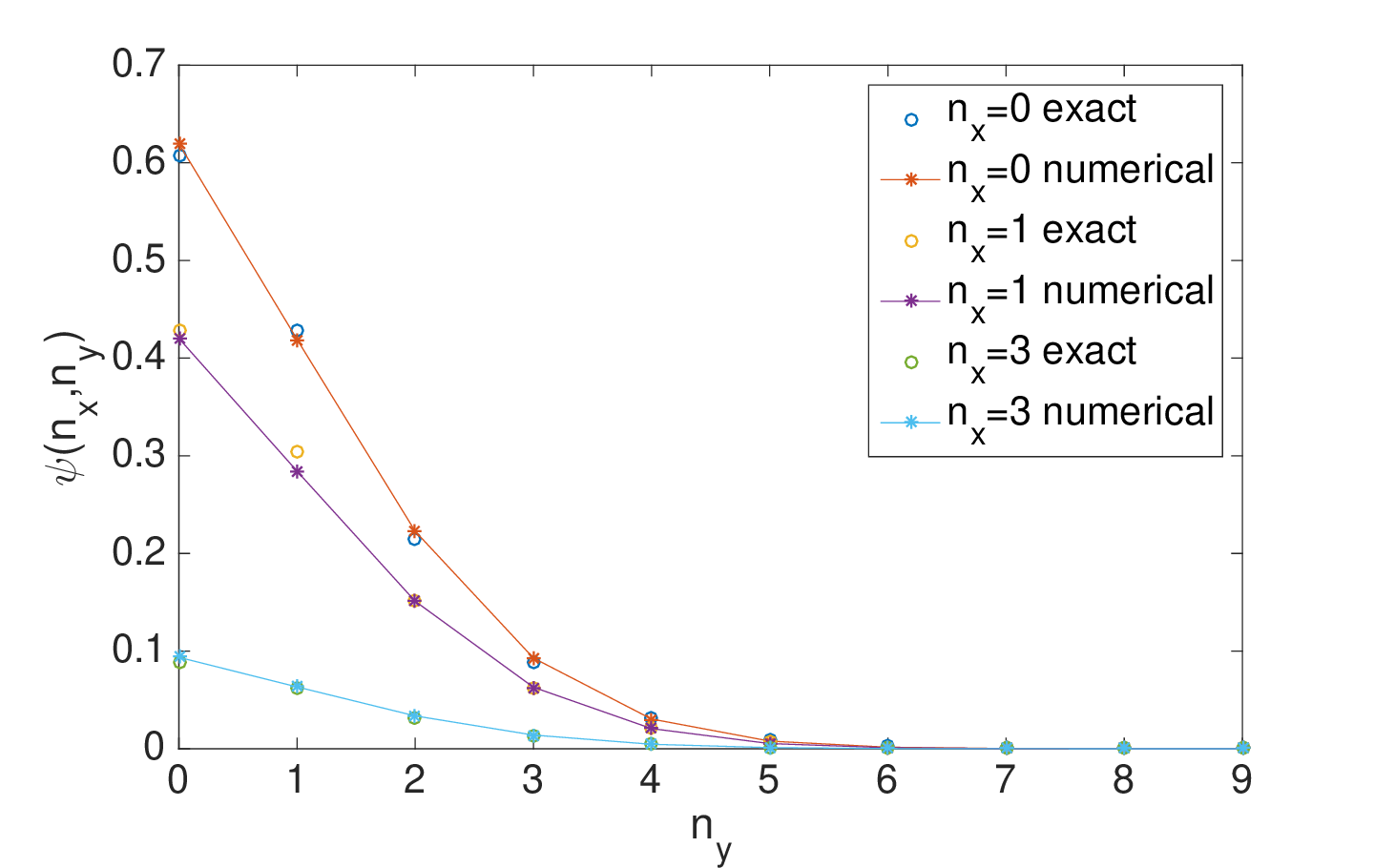}
	\caption{$\psi(n_x, n_y)$ as a function of $n_y$ at different $n_x$ with $E_x=1.0$,$E_y=1.0$. Circles represent theoretical values and asterisk represents the values with RBF network. In this figure, $\psi(n_x, n_y)$ is normalized.}
\label{fig:fig8}
\end{figure}

Table \ref{table1} lists a sample of the relation  between $nmax$ and the VMC energy at $E_x=4.0,E_y=2.0$. We can see that in this example the accuracy of the results improve with $nmax$.

\begin{table}[!htb] 
\caption {The relation between $nmax$ and the  VMC energy at $E_x=4.0,E_y=2.0$. VMC energy converges at $-8.99571\pm 0.00627$. Exact value is $9$. }
\begin{center}
 
	\begin{tabular}{| c | c |}
		\hline
		nmax & VMC energy \\ \hline
		3 & -6.28397 \\ \hline
		4 & -7.80747 \\ \hline
		5 & -8.02855 \\ \hline
		10 & -8.71073 \\ \hline
		20 & -8.90894 \\ \hline
		40 & -8.99571 \\ \hline

		\end{tabular}
\label{table1}

\bigskip

\end{center}
\end{table}

Figure. \ref{fig:fig8} shows  $\psi(n_x, n_y)$ as a function of $n_x$ and $n_x$ under different $E=(1.0,1.0)$. We can see that numerical results agree well with exact results.

\subsection{Particle in a box}

Another example that is benchmarked is a particle in a box with perturbation.

Consider the Hamiltonian

 \begin{equation}
H=\dfrac{\hat{p}^2}{2}+V(x)+a\hat{x}=H_0+a\hat{x},
\end{equation}

with $V(x)=0$ when $0<x<1$ and $V(x)=\infty$ when $x$ takes other values. $a\hat{x}$ is a linear potential defined on $0<x<1$ with $a$ as a parameter.

In natural units, the ground state energy of $H_0$ is $\frac{\pi^2}{2}=4.9348$. The first order perturbation theory correction for the ground state energy is $a/2$. The second order perturbation will give a correction of $-0.002194 a^2$.

A radial basis neural network VMC simulation can be similarly carried out.  As always we choose the basis to be the eigenbasis of $H_0$. 50000 samples are used. Ten hidden neurons ( $m=10$ ) are chosen in our calculation . $nmax$ is set at $20$. The learning rates are set at $0.01$. The matrix element in the local energy can be calculated as
 
 \begin{equation}
\langle  n_1|a x| n_2 \rangle=a\dfrac{4[(-1)^{n_1+n_2}-1]n_1n_2}{(n_1-n_2)^2(n_1+n_2)^2\pi^2},
\end{equation}
when $n_1\neq n_2$. And
 
  \begin{equation}
\langle  n_1|a x| n_2 \rangle=0.5a,
\end{equation}
when $n_1= n_2$.

\begin{figure}[!htb]
	\includegraphics[width=90mm]{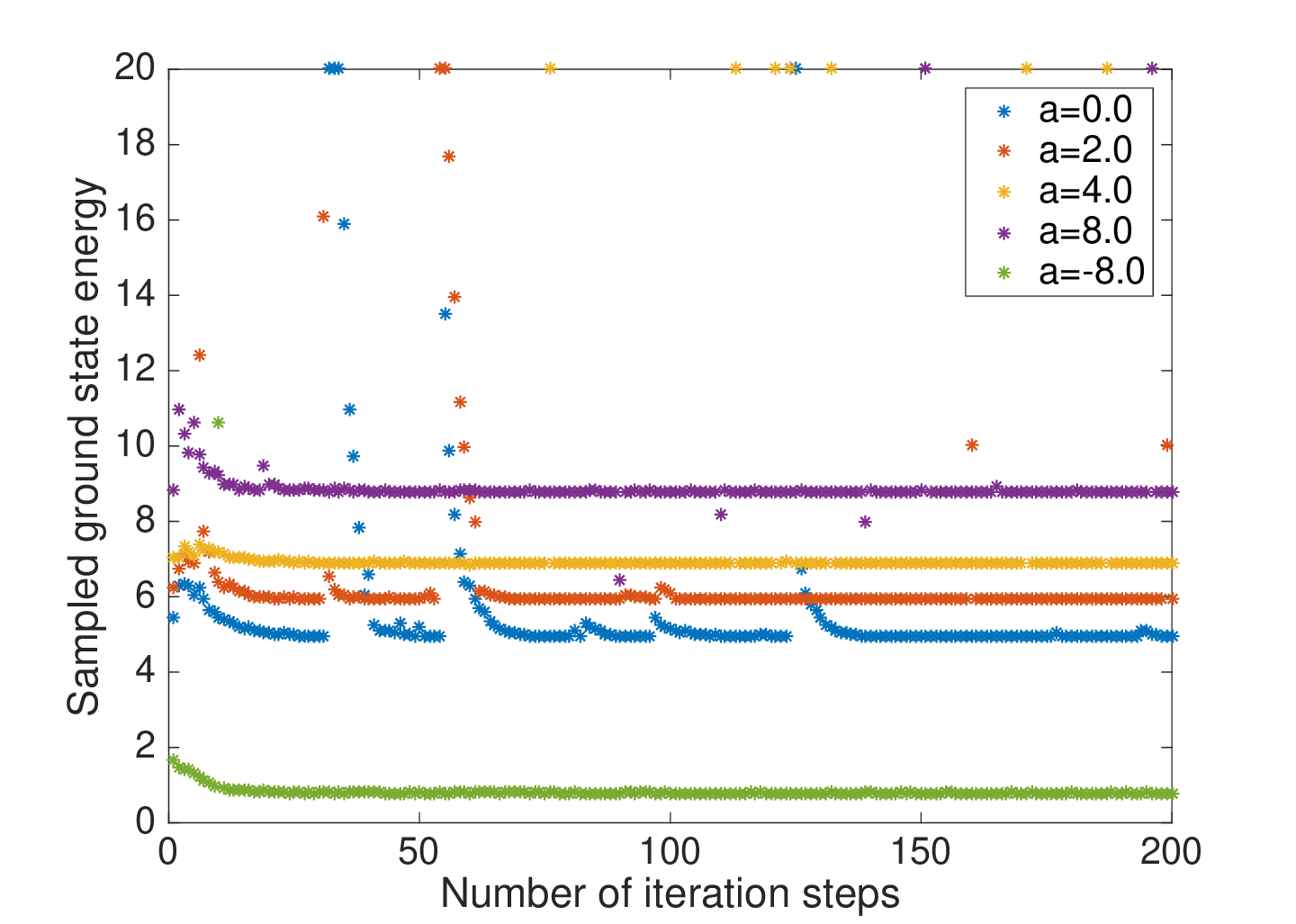}
	\caption{Minimization of the ground state energy of $H$ at $a=0.0,2.0,4.0,8.0,-8.0$.}
\label{fig:fig9}
\end{figure}

In Figure \ref{fig:fig9}, The convergence VMC ground state at different parameters is illustrated. Intermediate points that have a value which is larger than $20$ are set at 20 to maintain the scale of this graph. Notice that we get more spikes during the iteration when $a$ is small. The heights of the spikes decrease if smaller the learning rates are used.

Table \ref{table2} compares the result using an RBF network VMC, theoretical results up to second-order perturbation theory, and exact results. The exact ground state energy values are calculated using Mathematica. We can see that VMC performs much better than first-order perturbation theory and converge to the ground state energy that is very close to the theoretical ground state energy.

\begin{table}[!htb] 
\caption {Comparison between exact values, perturbation results and numerical VMC energy at different $a$. }
\begin{center}
 
	\begin{tabular}{| c | c | c | c | c |}
		\hline
		a & 1st order &  2nd order  & VMC energy & exact value\\ \hline
		0.0 & 4.9348 & 4.9348 & 4.9348$\pm$ 0.0001 & 4.93481 \\ \hline
		2.0 & 5.9348 & 5.9260 & 5.9260 $\pm$ 0.0001& 5.92603\\ \hline
		4.0 & 6.9348 & 6.8997 & 6.8998 $\pm$  0.0001 & 6.89974\\ \hline
		8.0 & 8.9348 & 8.7944 & 8.7960 $\pm$  0.0003& 8.79508\\ \hline
		-8.0 & 0.9348 & 0.7944 & 0.7950 $\pm$  0.0003& 0.795078\\ \hline

		\end{tabular}
\label{table2}

\bigskip

\end{center}
\end{table}

\subsection{Neural network as a Hermitian matrix lowest eigenvalue solver}

So far the examples that are benchmarked can all be solved by perturbation theory. Can neural network VMC method have a wider application than the perturbation theory? In this part, we will illustrate the possibility of using an RBF network VMC method to solve for the smallest eigenvalue of a Hermitian matrix. This problem is non-perturbative and purely mathematical, and our result implies that neural network VMC can have much broader scope than perturbation method.

Consider an $n \times n$ Hermitian matrix $H$. The eigenvector  that corresponds to the lowest energy is an $n$ dimensional vector.

We can write this eigenvector as

 \begin{equation}
\vec{x}=\sum_{i=1}^{n}\psi(i)\hat{i},
\end{equation}

and any vector in this finite vector space can be written in this form.

Define the objective function to be
 \begin{equation}
E=\vec{x}^* H \vec{x}.
\end{equation}

Then the smallest value of E corresponds to the lowest eigenvalue of $H$. And our goal is to find a set of parameters in neural network $\psi$ that minimize $E$. 

We can convert the matrix multiplication in $E$ into a discrete sum, which can be evaluated using the Metropolis algorithm. Instead of the energy eigenbasis, in this situation, we can choose our configuration space to be $n$ points, where $n$ is the dimension of vector $\vec{x}$, and the trial move would be from basis $\hat{i}$ to $\hat{i'}$. Therefore we can use the same VMC technique to minimize $E$.

Our previous examples can be essentially understood in this way since our Hamiltonians are truncated to a finite dimensional matrix.

To give a concrete implementation of this idea, we consider a matrix

 \begin{equation}
H(d)_{pq}=1/p+1/q.
\end{equation}

Here $H(d)$ is a $d \times d$ dimensional matrix. $p$, $q$ are the label for $H(d)$. And the matrix element on the $p$-th row and   $q$-th column equals $1/p+1/q$.

We use the RBF network ansatz to calculate the lowest eigenvalue of $H(d)$. Hidden neuron numbers are set at $20$. 50000 samples are chosen. Iteration undergoes $300$ steps and learning rate is  $0.01$. Table \ref{table3} shows the result of our VMC simulation.
\begin{table}[!htb] 
\caption {VMC results of the lowest eigenvalue of $H(d)$.}
\begin{center}
 
	\begin{tabular}{| c | c | c |}
		\hline
		d & exact value  &  VMC result \\ \hline
		2 & -0.0811 & -0.0811 $\pm$  0.0000 \\ \hline
		3 & -0.1874 & -0.1873 $\pm$  0.0002 \\ \hline
		5 & -0.4219 & -0.4220  $\pm$  0.0008 \\ \hline
		10 & -1.008 & -1.008 $\pm$  0.0008  \\ \hline

		\end{tabular}
\label{table3}

\bigskip

\end{center}
\end{table}

Our optimized neural network also yields the eigenvector that corresponds to the lowest eigenvalue. The components can be acquired by plugging in $i$ into $\psi(i)$. For example, when $n=10$, VMC gives a eigenvector $\overrightarrow{V}$, which is  (0.6851,0.1174,-0.0711,-0.1646,-0.2200,-0.2562,-0.2813,-0.2994,-0.3127,-0.3226), while the exact vector $\overrightarrow{V_0}$ is (0.6807,0.1194,-0.0677,-0.1613,-0.2174,-0.2548,-0.2816,-0.3016,-0.3172,-0.3297).
The Euclidean norm of the error $d=|\overrightarrow{V}-\overrightarrow{V_0}|=1.1\times 10^{-2}$.

We also calculate the relation between the accuracy and $m$ (the number of neurons in the hidden layer).  For $d=10$, the variational energy is $-0.0811,-0.0811,-0.9943,-1.0002$ for $m=5,10,15,20$ respectively. 

\textbf{Caveat}: The learning rate depends on the number of hidden neurons, and it has to be set by trial and error. We also have to point out that when $d>10$, the VMC optimization procedure may converge slowly or fail to converge. The stability also depends on forms of $H$. For some large ill-conditioned matrices, it is expected that the random sampling process will not capture all the matrix elements and lead to inaccurate results. 

\section{Discussion} \label{sec4}

Is it possible to use an RBF network with continuous variables as the variational wavefunction? This is possible for some certain Hamiltonians. For example, we can use an RBF network with a Gaussian basis as the variation wavefunction for the ground state of a harmonic oscillator. Based on our test, although this ansatz works perfectly for the harmonic oscillator, the iteration may not converge to the correct ground state when applied to other models. This test is trivial for the harmonic oscillator since its ground state is intrinsically a Gaussian function. For wavefunctions with continuous variables, the Kato's cusp condition\cite{CPA:CPA3160100201} poses strong constraints on the mathematical form of the wavefunction. A wavefunction that does not satisfy this condition will result in strong numerical instability in the VMC calculation.

How is this approach useful? This approach provides a new way to find the ground state energy of a quantum system. Compared with traditional variational Monte Carlo simulation, this method does not require choosing a specific wavefunction from our intuition.  Does this method depend on choosing a basis $|n\rangle$?  The example on the diagonalization of a Hermitian matrix illustrates that it doesn't depend on it as well, although a good basis may improve the accuracy and stability.

One advantage of ANN-based VMC is that the code is easy to modularize. When programming, we can write the modules for a neural network, Hamiltonian, and optimization separately. For the same Hamiltonian, we can also compare the representation power of different neural networks and different optimization methods. This greatly reduces programming difficulties and improves accuracy.

A potential issue with the neural network VMC method is that the optimization algorithm may fail to find the global minimum of the objective function. This is a common issue in machine learning methods. We see that the stochastic reconfiguration may not work  well enough that it could find the smallest eigenvalue of a matrix of arbitrarily large dimension.  Therefore, finding a stable algorithm or stable neural network mathematical form for the VMC optimization should be a crucial task. If successful, the neural network VMC method may give numerical conclusions to many unsolved problems in quantum physics.

Based on the above points, one important research direction is to develop more efficient VMC optimization algorithms. Another interesting direction is to discuss the representation power of different neural networks since there are a variety of neural networks developed by the machine learning community. For example, one interesting problem is the representation power of a continuous restricted Boltzmann machine\cite{10.1007/3-540-46084-5_58}.  With a Gaussian activation function, a continuous restricted Boltzmann machine has some similarities with the RBF network ansatz discussed in this paper. It is promising to provide more accurate results due to the elegant mathematical structure of the restricted Boltzmann machine.

\bigskip
\section{Conclusion} 

In this article, RBF networks are used as the variational wavefunction for quantum systems, and VMC calculations are carried out. For the examples that are examined, the VMC results agree well with theoretical predictions. Furthermore, it is possible to use the VMC method to calculate the lowest eigenvalue of a Hermitian matrix.

\section*{Acknowledgement}

Great thanks should be given to Dr. Yuan-Ming Lu for his helpful discussions and comments. I also want to thank the Ohio State University Physics Department for supporting my study. This work is supported by the startup funds of Dr. Yuan-Ming Lu at the Ohio State University.

\section*{References}	

\bibliography{mybibfile.bib}

\end{document}